\documentclass[12pt]{article}
\usepackage[a4paper, total={7in, 10in}]{geometry}
\usepackage{setspace}
\usepackage{amsmath,amssymb,graphicx,bm}
\renewcommand\vec[1]{\bm{#1}}% droit
\newcommand\tsr[1]{\mathsf{#1}}
\usepackage{tcolorbox}
\usepackage{color}
\usepackage{cancel}
\usepackage{placeins}
\usepackage{titlesec}
\usepackage{fancyhdr}
\usepackage{fancybox}
\usepackage{mathtools}
\usepackage{comment}
\titleformat{\section}[hang]{\Large\sffamily\bfseries}%
{\rlap{\thesection}}{2em}{}
\titleformat{\subsection}[hang]{\large\sffamily\bfseries}%
{\rlap{\thesubsection}}{3em}{}

\usepackage{amsthm}
\usepackage[titletoc]{appendix}
\usepackage[hidelinks]{hyperref}

\DeclareMathAlphabet{\mathsfbi}{OT1}{cmss}{bx}{n}
\DeclareMathVersion{sfletters}
\SetSymbolFont{letters}{sfletters}{OML}{ntxsfmi}{b}{it}

\makeatletter
\newcommand{\mathbfsbilow}[1]{%
  \text{\mathversion{sfletters}$\m@th#1$}%
}

\DeclareRobustCommand{\tsr}[1]{%
  \begingroup
  \ifcat\noexpand #1\relax
    \edef\greek@test{\detokenize{#1}}%
    \edef\greek@test{\expandafter\@cdr\greek@test\@nil}%
    \edef\greek@test{\expandafter\@car\greek@test\@nil}%
    \edef\x{\the\lccode\expandafter`\greek@test}%
    \edef\y{\number\expandafter`\greek@test}%
    \ifnum\x=\y\relax
      \mathbfsbilow{#1}%
    \else
      \mathsfbi{#1}%
    \fi
  \else
    \mathsfbi{#1}%
  \fi
  \endgroup
}
\makeatother

\def\u{\vec{u}}
\def\x{\vec{x}}
\def\X{\vec{X}}
\def\y{\vec{y}}
\def\ub{\bar{\vec{u}}}
\def\pb{\bar{p}}
\def\S{\tsr{S}}
\def\ttau{\tsr{\tau}}
\def\tS{\tsr{S}}
\def\tOmega{\tsr{\Omega}}
\newcommand\transp{^\textsf{T}}
\renewcommand{\hat}{\widehat}
\renewcommand{\bar}{\overline}

\def\Ebb{\mathbb{E}}

\def\Rbb{\mathbb{R}}

\newcommand{\Dpu}[2]{\frac{\partial #1}{\partial #2}}

\DeclareMathOperator{\gradd}{grad\kern-.5em{grad}}

\DeclareMathOperator{\tr}{tr}

\renewcommand{\geq}{\geqslant}
\renewcommand{\leq}{\leqslant}
\let\oldtimes\times
\renewcommand{\times}{\!\oldtimes\!}

\newtheorem{theorem}{Theorem}

\newcommand{\smallO}[1]{\ensuremath{\mathop{}\mathopen{}o\mathopen{}\left(#1\right)}}

\theoremstyle{definition}

\newtheorem{remark}{Remark}
\newtheorem*{definition*}{Definition}

\title{Vorticity-dependent and symmetry-preserving LES models}
\excludecomment{appendix}
\begin{document}

\author{Oscar Cosserat\footnote{Göttingen Mathematisches Institut, Georg-August-Universität Göttingen, Office 021, Hauptgebaüde, Bunsenstraße 3-5, 37073 Göttingen - Germany. Oscar.Cosserat@mathematik.uni-geottingen.de}, Dina Razafindralandy\footnote{Laboratoire des Sciences de l’Ingénieur pour l’Environnement (LaSIE) - UMR CNRS 7356 -- La Rochelle Université, Pôle Sciences et Technologie, Avenue Michel Crépeau, F17042 La Rochelle
Cedex 1, France. Dina.razafindralandy@univ-lr.fr} \, and Can Selçuk\footnote{Univ Gustave Eiffel, Univ Paris Est Creteil, CNRS, UMR 8208, MSME, F-77454 Marne-la-Vallée, France. can.selcuk@univ-eiffel.fr}}
\date{}

\maketitle

\begin{abstract}
Within the Large Eddy Simulation framework, we propose a methodology based on the Lie theory to derive symmetry-preserving turbulence models. We apply this methodology to the incompressible Navier-Stokes equations. These models explicitly depend on both the filtered strain-rate tensor and the filtered vorticity tensor. Particular emphasis is placed on models that additionally ensure stability.  
\end{abstract}

\textbf{\emph{Keywords}}: Large Eddy Simulation, Lie group theory,  invariance theory, vorticity tensors, symmetry-preserving turbulence models

\subsubsection*{Acknowledgements}
We wish to thank Aziz Hamdouni and Maurice Rossi for interesting and motivating insights. We are also thankful to the GDR CNRS $n^\circ 2043$ "G\'eom\'etrie Diff\'erentielle et M\'ecanique" for stimulating discussions. The first author acknowledges the Deutsche Forschungsgemeinschaft RTG 2491 for its financial support.

\section{Introduction}

Playing a critical role in numerous natural and engineering processes, such as atmospheric flows, aerodynamics, combustion in engines or biomedical applications, turbulence is one of the most complex phenomena in fluid dynamics. Despite this widespread presence, understanding and accurately predicting turbulent flows remain a significant challenge due to the inherently chaotic, multi-scale nature of turbulence. Various methods have been developed to simulate and analyze turbulent flows. Among them are Direct Numerical Simulation (DNS) and Reynolds-Averaged Navier-Stokes (RANS), which have their limitations; DNS is computationally prohibitive for high Reynolds number flows, while RANS often lacks the fidelity required to capture the intricate dynamics of turbulent structures.

Lying between DNS and RANS, Large-Eddy Simulation (LES) addresses these limitations by resolving the large, energy-carrying eddies directly, while modeling the smaller, universal scales. LES not only reduces the computational cost compared to DNS but also provides more detailed and accurate flow predictions than RANS. The fundamental principle of LES lies in the spatial filtering of the Navier-Stokes equations, which separates the large scales from the small scales. The large scales, which are responsible for the majority of the turbulent kinetic energy and are highly dependent on the flow geometry, are resolved explicitly. In contrast, the small scales, which are more universal and isotropic, are modeled using subgrid-scale (SGS) models. The accuracy and efficiency of LES heavily depend on the choice of the SGS model, which must adequately represent the effects of the unresolved scales on the resolved ones.

Various approaches have been employed over decades to develop SGS models \cite{Sag06}. In this article, we posit that a model must exhibit a certain level of universality. In particular, as noted by Speziale \cite{Speziale1979} and followed by Oberlack \cite{oberlack01} and Razafindralandy et al. \cite{razafindralandy06a,tcfd07,physicaa12}, preserving the symmetry group of the underlying equations is crucial, since the symmetries reflect fundamental physical properties of the flow. 

Symmetry groups, particularly Lie groups, play a fundamental role in mathematical physics. They provide a powerful framework for understanding and solving complex physical systems. For example, Noether's theorem establishes a deep connection between symmetries and conservation laws \cite{Olver1993,ibragimov,bluman10}. Symmetries also enable to study integrability and find exact solutions to partial differential equations. In fluid mechanics, these solutions can serve as model flows in simplified conditions \cite{grassi00,dcds24}. Moreover, symmetries are the foundation of gauge theories. In the framework of computational physics, symmetry group theory has been used to build robust numerical schemes \cite{kim04,chhay10b,hoarau07}. Though, as shown in \cite{razafindralandy06a,physicaa12} almost all SGS models in the literature break the symmetries of the Navier-Stokes equations.

Symmetry-preserving SGS models have been proposed and numerically validated in \cite{razafindralandy06a} for isothermal fluid flows and in \cite{physicaa12} in the anisothermal case. In these papers, the SGS model depends on the resolved velocity only through the symmetric part of its gradient, i.e. the strain-rate tensor, for simplicity reasons. However, most of the models in the literature also incorporate the skew-symmetric part of the velocity gradient, namely the vorticity tensor. In the present article, we propose symmetry-preserving SGS models which contains not only the strain-rate tensor but also the vorticity tensor. As a consequence, the model contains not only the deformation part but also the rotation part of the resolved flow. 

As will be seen, the symmetry approach produces a wide class of symmetry-preserving SGS models. One way to restrict this class is to select models which ensure that the velocity remains bounded over time. This can be achieved by requiring the models to derive from a scalar potential \cite{razafindralandy06a}, like the viscous strain tensor, and ensuring that the total (viscous and subgrid) dissipation is positive.

This article is organized as follows. Section \ref{sec:preli} is dedicated to some reminders. First, we recall the Lie symmetry groups of the Navier-Stokes equations with hints on how they were computed. The Large-Eddy Simulation approach is then presented. Our main result is given in sections \ref{sec:inv_tensor1} and \ref{sec:inv_tensor2} where a class of symmetry-preserving SGS models is developed. At the end of section \ref{sec:inv_tensor2}, a necessary and sufficient condition for a polynomial subgrid tensor depending on strain-rate and vorticity tensors to be invariant is given. As a secondary result, a sufficient condition on such invariant tensors to produce positive dissipation is given in section \ref{sec:pot_tensor}. Section \ref{sec:ex_and_comp} is devoted to some discussion and comparison with some existing models.

\section{Preliminaries on the incompressible Navier-Stokes equations}\label{sec:preli}

Consider the incompressible Navier-Stokes equations
\begin{align}\label{eq:NS}
  &  \frac{\partial\vec{u}}{\partial t} + \vec{\nabla}\cdot\left(\vec{u}\otimes \vec{u} \right) - 2 \nu \vec{\nabla}\cdot \S + \frac{1}{\rho}\vec{\nabla}p = 0,\nonumber\\ 
  &\vec{\nabla}\cdot \vec{u} = 0,
\end{align}
where $\vec{u}=(u_1,u_2,u_3)$ is the velocity, $p$ is the pressure, $\S = \frac{1}{2}(\vec{\nabla} \vec{u} + \vec{\nabla} \vec{u}\transp)$ is the strain-rate tensor, $\nu$ the kinematic viscosity and $\rho$ the volumetric mass density of the fluid. Equations \eqref{eq:NS} are known to be invariant under a set of point transformations, that we recall briefly for the sake of completeness.

\subsection{Lie symmetries of the incompressible Navier-Stokes equations}\label{sec:symmetries}

Consider first a generic differential equation 
\begin{equation}
  \Ebb(\vec y) = 0
\end{equation}
where $\vec y$ is the list of independent and dependent variables. Consider a set of transformations
\begin{equation}
  T_{\epsilon} \colon \y \longmapsto T_{\epsilon}(\y)= \hat{\y} 
\end{equation}
depending smothly on a parameter $\epsilon\in \mathbb R$ and which forms a group. This set is called a symmetry group of \eqref{eq:NS} if, for each $\epsilon$,
\begin{equation}\label{symmetry}
    \Ebb(\hat{\vec y}) = 0 \qquad\iff\qquad \Ebb(\vec y) = 0.
\end{equation}
This condition means that each element of the group transforms any solution into another one.

It is generally not easy to find all the symmetry groups of an equation from Definition \eqref{symmetry}. Instead, one uses an infinitesimal condition. To this aim, assume that the group is additive and that $T_{\epsilon}$ is the identity map when $\epsilon=0$. Define the vector field
\begin{equation}
  \X=\sum_{i}X_i\frac{\partial}{\partial y_i} \qquad\text{where}\qquad X_i=\frac{\partial \hat{y}_i}{\partial \epsilon}_{|\epsilon = 0}
\end{equation}
which describes the action of the group around the identity. $\X$ is called the infinitesimal generator of the group since $T_{\epsilon}$ is the flow of $\X$. Under some regularity hypothesis on the equation, definition~\eqref{symmetry} is equivalent to the following infinitesimal condition \cite{bluman10, Olver1993}:
\begin{equation}\label{infiniteismal}
  \operatorname{pr}\X\cdot\,\Ebb\ (\vec y)=0\qquad \text{whenever }\qquad\Ebb(\vec y)=0
\end{equation}
where $\operatorname{pr}\X$ is a prolongation of $\X$ which incorporates the variations at the identity of derivatives present in the equation.

For the Navier-Stokes equations \eqref{eq:NS}, $\y=(t,\x,\u,p)$ where $\x=(x_1,x_2,x_3)$ is the spatial variable. Solving condition \eqref{infiniteismal} yields the exhaustive list of infinitesimal symmetry generators of \eqref{eq:NS}. Upon integration, these generators provide the corresponding symmetry groups. The infinitesimal generators and their associated groups are enumerated below (see also \cite{Pukhnachov1972,ibragimov}).

\begin{itemize}
  \item The group of time translations 
	\begin{equation*}
	  G_t=\{\left(t,\vec{x},\vec{u},p \right)  \longmapsto \left(t + \epsilon,\vec{x},\vec{u},p \right),\ \epsilon\in\mathbb R\}
  \end{equation*}
  which admits $\displaystyle \frac{\partial }{\partial t}$ as infinitesimal generator.
\item The generalized Galilean transformation group:
\begin{center}
  $Gal: \{ \left(t,\vec{x},\vec{u},p \right)  \rightarrow \left(t ,\vec{x}+\vec{\alpha}(t),\vec{u} + \vec{\dot{\alpha}}(t),p -\rho  \vec{\ddot{\alpha}}(t)\cdot \vec{x} \right),\ \vec{\alpha}\in C^2(\Rbb,\Rbb^3)\}$
\end{center}
generated by the vector fields
\begin{equation*}
  \alpha_i(t) \frac{\partial }{\partial x_i} + \dot{\alpha}_i(t) \frac{\partial}{\partial u_i} - \rho x_i \ddot{\alpha}_i(t) \frac{\partial }{\partial p}, \quad 1 \leq i \leq 3
\end{equation*}
\item The group $SO(3)$ of rotations acting on $\x$ and $\vec{u}$
\begin{center}
  $SO(3)=\{\left(t,\vec{x},\vec{u},p \right)  \longmapsto\left(t ,\tsr{R}\vec{x},\tsr{R}\vec{u},p \right),\ \tsr R\,\tsr R\transp=\tsr{I}\tsr{d},\ \operatorname{det}\tsr R=1\}$
\end{center}
associated to the three infinitesimal generators
\begin{equation*}
  x_j \frac{\partial}{\partial x_i} - x_i \frac{\partial }{\partial x_j} + u_j \frac{\partial }{\partial u_i} - u_i \frac{\partial }{\partial u_j}, \quad i = 1,2, \quad i \leq j \leq 3.
\end{equation*}
\item The pressure translations
\begin{equation*}
  G_p=\{\left(t,\vec{x},\vec{u},p \right)  \longmapsto \left(t ,\vec{x},\vec{u},p + \xi(t) \right),\ \xi\in C^0(\mathbb R,\mathbb R)\}
\end{equation*}
admitting the infinitesimal generator $\displaystyle \xi(t) \frac{\partial}{\partial p}$.
\item The group of scale transformations
\begin{center}
	$G_s=\{\left(t,\vec{x},\vec{u},p \right)  \longmapsto \left(e^{2 \epsilon}t ,e^{\epsilon}\vec{x}, e^{- \epsilon} \vec{u}, e^{- 2 \epsilon}p  \right)\},$
\end{center}
obtained from the infinitesimal generator
\begin{equation*}
2t \frac{\partial}{\partial t} + \sum\limits_{j=1}^3 x_j \frac{\partial}{\partial x_j} - \sum\limits_{j = 1}^3 u_j \frac{\partial }{\partial u_j} - 2p \frac{\partial}{\partial p}.
\end{equation*}

\end{itemize}

$G_t$ and $SO(3)$ are a symmetry groups of \eqref{eq:NS} since these equations do not depend on the choice of the origin of time and on the orientation of the (space) frame. Next, the equations are invariant not only under a change of the origin of the frame or under a Galilean boost but also under any time-dependent shift of the frame, provided that the velocity and pressure are corrected suitably as in the expression of the elements of $Gal$. The invariance under $G_p$ reflects the fact that the pressure appears in the equations solely through its gradient. Lastly, invariance under $G_s$ illustrates the way in which velocity and pressure vary in response to a particular space-time scaling. Groups $G_t$ and $G_s$ are one-dimensional Lie groups and $SO(3)$ is a three-dimensional Lie group whereas $Gal$ and $G_p$ are infinite dimensional groups.

It is interesting to note that the Euler equations of an ideal fluid possess one more symmetry than those listed above. This additional symmetry corresponds to the following scaling transformation for the Navier-Stokes equations
\begin{equation}
  (t,\x,\u,p,\nu)\longmapsto (t,e^a\x,e^a\u,e^{2a}p,e^{2a}\nu).
  \label{eq:transformation_viscosity}
\end{equation}
Although this transformation keeps equations \eqref{eq:NS} invariant, it is not strictly speaking a symmetry because it involves viscosity. For this reason, it will not be considered further in the subsequent discussion.

Another known symmetry of the Navier-Stokes equations is the material indifference \cite{speziale98,Pope_2000}. But since this symmetry exists only under a rather drastic condition (for instance, in the limit of bidimensional flow), it will not be used herein.

\subsection{Large Eddy Simulation (LES)}\label{sec:LES}

The LES approach relies on a reduction of the computational cost by removing small scales of $\u$ and $p$: those smaller than the typical grid size. The resolved scales (those which are bigger than the typical grid size and effectively computed) may be seen as filtered versions of $\u$ and $p$.

The reduction process may be understood as a filtering with an implicit kernel $K$. The resolved velocity and pressure can then be defined as follows
\[
\ub=K\star \u \qquad \text{and}\qquad \pb=K\star p
\]
where $\star$ is a convolution operation.

Applying the filtering to the incompressible Navier-Stokes equations \eqref{eq:NS}, we obtain the equations of  $\ub$ and $\pb$:
\begin{align}\label{eq:filtered_NS}
  &  \frac{\partial\bar{\vec{u}}}{\partial t} + \vec{\nabla}\cdot\left(\bar{\vec{u}}\otimes \bar{\vec{u}} \right)  +\frac{1}{\rho}\vec{\nabla}\bar{p}- 2 \nu \vec{\nabla}\cdot \bar{\S} + \vec{\nabla} \cdot \ttau=0,\nonumber\\ 
  &\vec{\nabla}\cdot \bar{\vec{u}} = 0,
\end{align}
where $\bar{\S} = \frac{\nabla \bar{\vec{u}} + \nabla \bar{\vec{u}}\transp}{2}$ is the filtered strain-rate tensor and $\ttau = \overline{\vec{u}\otimes \vec{u}} - \bar{\vec{u}}\otimes \bar{\vec{u}}$ is the subgrid tensor. It must be modelled as a function of the resolved scales to close equations \eqref{eq:filtered_NS}. More specifically, $\ttau$ will be modelled as a function of the filtered strain-rate tensor $\bar{\tS}$ and the filtered vorticity tensor $\displaystyle\bar{\tOmega} = \frac{\nabla \bar{\vec{u}} - \nabla \bar{\vec{u}}\transp}{2}$.

We will say that a model of $\ttau$ is invariant under a symmetry group $G$ of equation \eqref{eq:NS} if $G$ is also a symmetry of the filtered equations \eqref{eq:filtered_NS} when applied to the filtered variables $(t,\x,\ub,p)$. Our primary goal is to propose models which are invariant under each of the previously described symmetry groups.

\emph{
In the sequel, we will only deal with filtered variables. Thus, we remove all the overbars $\bar{\,\centerdot\,}$ in filtered quantities. More precisely, from now on, $\u,$ $p$, $\tS$ and $\tOmega$ will denote respectively the filtered velocity, the filtered pressure, the filtered strain-rate tensor and the filtered vorticity tensor.
}

\section{Invariance under \texorpdfstring{$G_t$}{Gt}, \texorpdfstring{$Gal$}{Gal}, \texorpdfstring{$SO(3)$}{SO3} and \texorpdfstring{$G_p$}{Gp}}\label{sec:inv_tensor1}

In this section, we propose a subgrid tensor which is invariant under $G_t$, $Gal$, $SO(3)$ and $G_p$, postponing the invariance under the scale transformation group $G_s$ until the next section. To this aim, we use the invariant theory in \cite{boehler1987}. 

Before announcing our result, let us first introduce some notations. The deviatoric part of a 2-tensor $\tsr Q$ will be denoted $\tsr Q^d$, that is 
\[
\tsr Q^d=\tsr Q-\frac 13\tr \tsr Q.
\]

The commutator of two 2-tensors $\tsr P$ and $\tsr Q$ is denoted $[\tsr P,\tsr Q]$, and defined by
\[
[\tsr P,\tsr Q]=\tsr P\tsr Q-\tsr Q\tsr P.
\]

\begin{theorem}\label{thm:zheng}
Let $\ttau$ be the subgrid tensor with deviatoric part
\begin{equation}
	\ttau^d =  \alpha_1 \S + \alpha_2\,(\S^2)^d + \alpha_3\,(\tOmega^2)^d + \alpha_4\,\left(\tOmega\S\tOmega\right)^d +\alpha_5 \left[\S,\tOmega\right] + \alpha_6\left[\S^2,\tOmega\right] + \alpha_7\left[\tOmega\S\tOmega,\tOmega\right],
  \label{eq_tau_zheng}
\end{equation}
where $\alpha_1, ..., \alpha_7$ are abitrary scalar functions of the variables $\left(I_1,I_2,B_1,B_2,B_3,B_4 \right)$ defined by 
\begin{equation}\label{eq_dependencies}
  \begin{array}[]{lllll}
	I_1=\tr\left(\S^2\right),&&I_2 = \tr\left(\S^3\right),&&B_1 =\tr\left(\S^2\tOmega^2\right)\\[5pt]
	B_2 = \tr\left(\tOmega^2 \right),&&  B_3 = \tr\left(\S\tOmega^2 \right),&& B_4 = \tr\left(\vec{S^}2\tOmega^2\S\tOmega\right).
  \end{array}
\end{equation}
Then $\ttau$ is invariant under the symmetry groups $G_t$, $Gal$, $SO(3)$ and $G_p$.
Moreover, the representation (\ref{eq_tau_zheng}) is minimal.
\end{theorem}

\begin{proof}
The existence of the functional basis \eqref{eq_dependencies} and of the tensorial basis
\begin{equation}\label{eq:basis}
    (\tsr I\tsr d, \tS, \tOmega^2, \tS^2, \tS \tOmega \tS, [\tOmega, \tS], [\tS^2, \tOmega], [\tOmega \tS \tOmega, \tOmega])
\end{equation}
for an analytic isotropic tensor $\tau$ depending on $S$ and $\tOmega$ are shown in \cite{Zheng1993}. Their minimality are proven in \cite{Pennisi1987}. We still have to check the invariance of $\tau$ under the symmetry groups.

Since $\tau$ depends only on the first partial derivatives of the velocity, it is invariant under $G_t$, $Gal$ and $G_p$. Moreover, under an element of $SO(3)$, $\ttau$ transforms as follows
\begin{equation*}
    \hat{\ttau} = \tsr R \ttau \tsr R^{-1}.
\end{equation*}
where $\tsr R$ is an arbitrary rotation tensor. It is then straightforward to check that equations \eqref{eq:filtered_NS} are invariant under $SO(3)$.
\end{proof}

\begin{remark}
The question of the representation of tensors carrying invariance properties is now more than a century years old (\cite{Gordan1900, Hilbert1993}). Some of those old results have been rediscovered along the second half of the twentieth century and applied to mechanics. We track back this history there. A first method to compute a generating set with 11 elements for an isotropic tensor depending on a symmetric and an anti-symmetric tensor is given in \cite{rivlin1955}. The book \cite[Sec. 7.2.2]{Sagaut1998}, and \cite{lund1992} state this result without any proof. The article \cite{pope1975} also recovers a generating set with 11 elements, and disagrees furthermore with the invariants \eqref{eq_dependencies} by providing less invariants than the one we have here.
The equation (4.5) of \cite{smith1971} provides a smaller basis than the one of \cite{pope1975}. This smaller basis is the one we use here. As stated before, \cite{Pennisi1987} proved the minimality of this smaller basis by a direct way and closed the discussion on the minimality of this way of representing an invariant subgrid tensor depending on the strain and vorticity tensors $\tS$ and $\tOmega$.
\end{remark}

\section{Invariance under \texorpdfstring{$G_s$}{Gs}}\label{sec:inv_tensor2}

In this section, we restrict the class of models in theorem \ref{thm:zheng} so that the subgrid scale model $\ttau$ is also invariant under the symmetry group $G_s$ of the Navier-Stokes equations. This is done by suitably imposing the form of the arbitrary functions $\alpha_i$ which leads to another functional basis. The results are summarized in theorem \ref{thm:NS_invariant} which delivers a class of subgrid tensors depending on the strain and vorticity tensors, and which are invariant by all the symmetry groups of the incompressible Navier-Stokes equations.

\begin{theorem}\label{thm:NS_invariant}
A subgrid tensor of the form
\begin{equation}
  \begin{array}{lll}
	\ttau^d &=&  \displaystyle \alpha_1^0 \S + 
	\frac{\alpha_2^0}{|\S|} (\S^2)^d + 
	\frac{\alpha_3^0}{|\S|} (\tOmega^2)^d +
	\frac{\alpha_4^0}{|\S|^2}\left(\tOmega\S\tOmega\right)^d \\[20pt]
	&+& \displaystyle 
	\frac{\alpha_5^0}{|\S|} \left[\S,\tOmega\right] + 
	\frac{\alpha_6^0}{|\S|^2}\left[\S^2,\tOmega\right] + 
	\frac{\alpha_7^0}{|\S|^3}\left[\tOmega\S\tOmega,\tOmega\right]
  \end{array}
  \label{eq:tau_1erchgt}
\end{equation}
where $|\S|=\sqrt{\tr(\S^2)}$ and $\alpha_1^0,\dots,\alpha_7^0$ are arbitrary scalar functions of
\begin{equation}
  \begin{array}[]{lllllllll}
	\displaystyle v_1=\frac{\tr(\S^3)}{|\S|^3},&&
	\displaystyle v_2=\frac{\tr(\S^2\tOmega^2)}{|\S|^4},&&
	\displaystyle v_3=\frac{\tr(\tOmega^2)}{|\S|^2},\\[20pt]
	\displaystyle v_4=\frac{\tr(\S\tOmega^2)}{|\S|^3},&&
	\displaystyle v_5=\frac{\tr(\S^2\tOmega^2\S\tOmega)}{|\S|^6}
  \end{array}
  \label{eq:v1}
\end{equation}
is invariant under the symmetry groups $G_t$, $Gal$, $SO(3)$, $G_p$ and $G_s$ of the Navier-Stokes equations.
\end{theorem}

\begin{proof}
  We start with a subgrid tensor model $\ttau$ of the form \eqref{eq_tau_zheng}. Recall that $G_p$ consists of maps
\begin{equation}\label{eq:1er_chgt}
  (t,\x,\u,p) \mapsto (e^{2 \epsilon} t, e^{\epsilon} \x, e^{-\epsilon} \u, e^{-2 \epsilon} p)
\end{equation}
with $\epsilon\in\mathbb R$.
These maps (when acting on the filtered velocity and pressure) are symmetries of the filtered equations \eqref{eq:filtered_NS} if and only if the subgrid model $\ttau$ transforms as
\begin{equation}
    \hat \ttau = e^{-2 \epsilon} \ttau
\end{equation} for all $\epsilon \in \Rbb.$ A necessary and sufficient condition to fulfill this condition is that the functions $\alpha_i$'s scale as follows:
\begin{equation}\label{eq:alpha1}
    \alpha_{1}(I_1, I_2, B_1, B_2, B_3, B_4) = \alpha_{1}( e^{- 4 \epsilon} I_1, e^{- 6 \epsilon} I_2, e^{- 8 \epsilon} B_1, e^{- 4 \epsilon}B_2, e^{- 6 \epsilon} B_3, e^{- 12 \epsilon} B_4),
\end{equation}
\begin{equation}\label{eq:alpha2}
    e^{2 \epsilon} \alpha_i(I_1, I_2, B_1, B_2, B_3, B_4) = \alpha_i(e^{- 4 \epsilon} I_1, e^{- 6 \epsilon} I_2, e^{- 8 \epsilon} B_1, e^{- 4 \epsilon}B_2, e^{- 6 \epsilon} B_3, e^{- 12 \epsilon} B_4)
\end{equation}
for $i\in\{2,3,5\}$, 
\begin{equation}\label{eq:alpha4}
    e^{4 \epsilon}\alpha_i(I_1, I_2, B_1, B_2, B_3, B_4) = \alpha_i(e^{- 4 \epsilon} I_1, e^{- 6 \epsilon} I_2, e^{- 8 \epsilon} B_1, e^{- 4 \epsilon}B_2, e^{- 6 \epsilon} B_3, e^{- 12 \epsilon} B_4)
\end{equation}
for $i\in\{4,6\}$, and
\begin{equation}\label{eq:alpha7}
    e^{6 \epsilon} \alpha_{7}(I_1, I_2, B_1, B_2, B_3, B_4) = \alpha_{7}(e^{- 4 \epsilon} I_1, e^{- 6 \epsilon} I_2, e^{- 8 \epsilon} B_1, e^{- 4 \epsilon}B_2, e^{- 6 \epsilon} B_3, e^{- 12 \epsilon} B_4).
\end{equation}
Differentiating equations \eqref{eq:alpha1}-\eqref{eq:alpha7} with respect to $\epsilon$ and taking $\epsilon=0,$ the method of characteristics shows that there exists scalar functions $\alpha_i^0$ such that \cite{olver2013, strauss2007}
\begin{equation}
  \begin{array}{l}
	\alpha_{1}(I_1, I_2, B_1, B_2, B_3, B_4) = \alpha_{1}^0(v_1,v_2,v_3,v_4,v_5),\\[10pt]
	\alpha_i(I_1, I_2, B_1, B_2, B_3, B_4) = (I_1)^{-\frac{1}{2}} \alpha_i^0(v_1,v_2,v_3,v_4,v_5),\qquad i\in\{2,3,5\},\\[10pt]
	\alpha_i(I_1, I_2, B_1, B_2, B_3, B_4) = (I_1)^{-1} \alpha_i^0(v_1,v_2,v_3,v_4,v_5),\qquad i\in\{4,6\} \text{ and}\\[10pt]
	\alpha_{7}(I_1, I_2, B_1, B_2, B_3, B_4) = (I_1)^{-\frac{3}{2}} a_{7}^0(v_1,v_2,v_3,v_4,v_5).
  \end{array}
\end{equation}
where
\begin{equation}
  \begin{array}{lll}
	v_1 = I_2 (I_1)^{-\frac{3}{2}},\qquad&
	v_2 = B_1 (I_1)^{-2},\qquad&
	v_3 = B_2 (I_1)^{-1},\\[20pt]
	v_4 = B_3 (I_1)^{-\frac{3}{2}},&
	v_5 = B_4 (I_1)^{-3}
  \end{array}
\end{equation}
are invariants of the group $G_s$.
Using the definitions of $I_1,\ I_2,\ B_1,\ B_2$ and $B_3$, one obtains expressions \eqref{eq:v1}.
\end{proof}

\begin{remark}
  As shown in \cite{ejm07}, $v_1$ is bounded: 
  \[|v_1|\leq v^*=\frac{1}{3 \sqrt{6}}.\]
\end{remark}

\section{A class of invariant models deriving from a potential}\label{sec:pot_tensor}

In the present section, we seek for conditions on the models such that the filtered velocity remains bounded. This, in turn, ensures the stability of the model. As proven in \cite{ejm07}, a sufficient condition for this is that the total (viscous and subgrid) dissipation
\begin{equation}\label{eq:total_dissipation}
  \Phi_T = \tr( (\ttau_{visc} - \ttau) \S)
\end{equation}
is positive. In \eqref{eq:total_dissipation}, $\ttau_{visc}=2\nu\S$ is the filtered viscous strain tensor. It derives from a convex potential $\phi_{visc}$ such that
\begin{equation}\label{eq:tau_visc}
\ttau_{visc}=\frac{\partial\phi_{visc}}{\partial \S}\qquad\text{where}\qquad\phi_{visc}(\S,\tOmega)=\nu\tr\S^2.
\end{equation}
By analogy with $\ttau_{visc}$, we ask the SGS model $\ttau$ to derive also from a real scalar potential $\phi(\S,\tOmega)$ in subsection \ref{subsec:pot_tensor}. More precisely, we ask that $\ttau$ is the derivative of $\phi(\S,\tOmega)$ with respect to $\S$. Using convexity property, we then show how to ensure the positivity of $\Phi_T$ in subsection \ref{subsec:pot_tensor_positive}.

\subsection{A condition for the subgrid tensor to derive from a potential}\label{subsec:pot_tensor}

In the same way as equation \eqref{eq:tau_visc}, we say that the SGS model $\ttau$ derives from a potential $\phi(\S, \tOmega)$ if
\begin{equation}
    \ttau = \frac{\partial \phi}{\partial \S}.
\end{equation}

The following theorem provides a class of SGS models fulfilling invariant conditions of the previous sections and, in addition, deriving from a potential.

\begin{theorem}\label{thm:pot_inv}
Let $\ttau$ be a subgrid tensor such that its deviatoric part is
\begin{align}\label{eq:pot_inv}
  \ttau^d &= \left(2g - 3 v_1 \frac{\partial g}{\partial v_1}-  2 v_3 \frac{\partial g}{\partial v_3}- 3  v_4 \frac{\partial g}{\partial v_4}\right)\S\\[10pt] \nonumber
	&+ \frac3{|\S|} \frac{\partial g}{\partial v_1}(\S^2)^d
    +\frac1{|\S|} \frac{\partial g}{\partial v_4}(\tOmega^2)^d,\nonumber
\end{align}
where $g$ is a scalar function of $v_1, v_3, v_4.$ Then, $\ttau$ is invariant under the symmetry groups $G_t$, $Gal$, $SO(3)$ and $G_p$ and derives from a scalar potential.
\end{theorem}

\begin{proof}
We require the SGS tensor to derive from a potential $\phi:$
\begin{equation}\label{eq:pot_phi}
  \ttau^d = \left(\Dpu{\phi}{\S}\right)^d.
\end{equation}
We also require $\ttau$ to fulfill invariant properties. In order to apply Theorem \ref{thm:zheng}, we assume $\phi$ to be a function of the initial primitive variables $I_1, I_2, B_1, B_2, B_3, B_4.$ Out of the chain rule, Equation \eqref{eq:pot_phi} is equivalent to 
\begin{equation*}
  \ttau^d  = \left(\Dpu{\phi}{I_1}\Dpu{I_1}{\S} + \Dpu{\phi}{I_2}\Dpu{I_2}{\S} + \Dpu{\phi}{B_1}\Dpu{B_1}{\S} + \Dpu{\phi}{B_2}\Dpu{B_2}{\S} + \Dpu{\phi}{B_3}\Dpu{B_3}{\S} + \Dpu{\phi}{B_4}\Dpu{B_4}{\S}\right)^d
\end{equation*}
The derivatives with respect to $\S$ read
\begin{equation}
  \begin{array}[]{lll}\displaystyle 
    \frac{\partial I_1}{\partial\S} = 2\S\qquad&\displaystyle 
    \frac{\partial I_2}{\partial\S} = 3\S^2\qquad&\displaystyle 
	\frac{\partial B_1}{\partial\S} = \S\tOmega^2 + \tOmega^2 \S\\[20pt]\displaystyle 
    \frac{\partial B_2}{\partial\S} = 0&\displaystyle 
    \frac{\partial B_3}{\partial\S} = \tOmega^2
  \end{array}
\end{equation}
and
\begin{equation}
  \frac{\partial B_4}{\partial\S} =\frac12\left(\S \tOmega^2 \S \tOmega - \tOmega \S \tOmega^2 \S + \tOmega^2 \S \tOmega \S - \S \tOmega \S \tOmega^2 + \tOmega \S^2 \tOmega^2 - \tOmega^2 \S^2 \tOmega  \right)
  \label{eq:db4}
\end{equation}
To establish equation \eqref{eq:db4}, note that by Riesz representation theorem, $\frac{\partial B_4}{\partial \S}$ is the unique symmetric matrix $\tsr B$ such that
\begin{equation}
    B_4(\tsr S+\tsr H) = B_4(\tsr S) + \tr(\tsr B\tsr H) + \smallO{\tsr H}
\end{equation} 
for any symmetric matrix $\tsr H$.
From the expression of $B_4(\tsr S+\tsr H)$ and the cyclic property of the trace operator, it follows that
%Using the fact that $\text{Tr}(xy) = \text{Tr}(yx)$ for any matrices $x$ and $y$ and that $H$ is a symmetric matrix:
\begin{align}
  \tr(\tsr B\tsr H) = \tr\left( (\S\tsr H + \tsr H\tsr S) \tOmega^2 S \tOmega + \tsr S^2 \tOmega^2 \tsr H \tOmega \right) = \tr(\tsr C\tsr H) 
\end{align}
where 
\begin{equation}
  \tsr C=\S\tOmega^2\S\tOmega + \tOmega^2\S\tOmega\S + \tOmega\S^2\tOmega^2
\end{equation}
Note that $\tsr C$ is not the derivative of $B_4$ because it is not a symmetric matrix. However, since $\tsr H$ is symmetric, 
\[
\tr(\tsr B\tsr H) =  \tr\left(\frac{\tsr C+\tsr C\transp}2\tsr H\right),
\]
meaning that the derivative $\tsr B$ of $B_4$ is the symmetric part of $\tsr C$, which is displayed in \eqref{eq:db4}.

So far, we showed that equation \eqref{eq:pot_phi} is equivalent to
\begin{equation}
  \begin{array}{lll}
	\ttau^d &=&\displaystyle  2\frac{\partial \phi}{\partial I_1} \S
	+ 3\frac{\partial \phi}{\partial I_2}(\S^2)^d
	+ \frac{\partial \phi}{\partial B_1} (\S \tOmega^2 + \tOmega^2 \S)^d 
	+\frac{\partial \phi}{\partial B_3}(\tOmega^2)^d\\[15pt]
	&+&\displaystyle \frac{1}{2}\frac{\partial \phi}{\partial B_4}(\S \tOmega^2 \S \tOmega - \tOmega \S \tOmega^2 \S + \tOmega^2 \S \tOmega \S - \S \tOmega \S \tOmega^2 + \tOmega \S^2 \tOmega^2 - \tOmega^2 \S^2 \tOmega)^d.
  \end{array}
\label{eq:dem_taugrad5}
\end{equation}
Comparing expression \eqref{eq:dem_taugrad5} with \eqref{eq:tau_1erchgt}, we conclude that a model which is invariant under all the symmetry groups of the Navier-Stokes equations derives from a potential if
\begin{equation}
    \alpha_4^0 = \alpha_5^0 = \alpha_6^0 = \alpha_7^0 = 0
\end{equation}
and
\begin{subequations}\label{potential_strict_ns}
\begin{align}
\label{potential_strict_ns_prem} 
   \Dpu{\phi}{B_1} &= \Dpu{\phi}{B_4} = 0, \\  
      2\Dpu{\phi}{I_1} = \alpha_1^0; ~~~~ 3 \Dpu{\phi}{I_2} &=  I_1^{-1/2}\alpha_2^0; ~~~~  \Dpu{\phi}{B_3} = I_1^{-1/2}\alpha_3^0.
\label{potential_strict_ns_dern}
\end{align}
\end{subequations}
Conditions \eqref{potential_strict_ns_prem} and \eqref{potential_strict_ns_dern} imply that $\alpha_1^0$ does not depend on $v_2.$ Indeed,
\begin{equation*}
\frac{\partial \alpha_1^0}{\partial v_2} =  \frac{\partial \alpha_1^0}{\partial B_1}\frac{\partial B_1}{\partial v_2} = 2 I_1^2 \frac{\partial^2 \phi}{\partial I_1\partial B_1} = 0.
\end{equation*}
$\alpha_1^0$ does not depend on $v_5$ either since
\begin{equation*}
\frac{\partial \alpha_1^0}{\partial v_5} =  \frac{\partial \alpha_1^0}{\partial B_4}\frac{\partial B_4}{\partial v_5} = 2 I_1^3 \frac{\partial ^2\phi}{\partial I_1\partial B_4} = 0.
\end{equation*}
Through analogous reasoning, it can be shown that $\alpha_2^0$ and $\alpha_3^0$ are independent of both $v_2$  and $v_5.$
In summary,
\begin{equation}
  \alpha_1^0 = \alpha_1^0\left(v_1,v_3,v_4\right),\qquad\alpha_2^0 = \alpha_2^0\left(v_1,v_3,v_4\right),\qquad\alpha_3^0 = \alpha_3^0\left(v_1,v_3,v_4\right).
\end{equation}

To conclude, we use conditions \eqref{potential_strict_ns_dern} to reduce the number of arbitrary scalar functions from three ($\alpha_1^0, \alpha_2^0, \alpha_3^0$) to a single one, namely $g = g\left(v_1,v_3,v_5\right)$. 
First, let us compute $\frac{\partial \alpha_1^0}{\partial v_1}.$
\begin{align}
  \frac{\partial \alpha_1^0}{\partial v_1} &= 2\frac{\partial I_2}{\partial v_1} \frac{\partial^2 \phi}{\partial I_1 \partial I_2}=\frac{2}{3} I_1^{\frac{3}{2}} \frac{\partial }{\partial I_1}\left(I_1^{-\frac{1}{2}} \alpha_2^0\right)\\[10pt]
    &= -\frac{1}{3} \alpha_2^0 - v_1 \frac{\partial \alpha_2^0}{\partial v_1} - \frac{2}{3} v_3 \frac{\partial \alpha_2^0}{\partial v_3} - v_4 \frac{\partial \alpha_2^0}{\partial v_4}.\label{eq:alpha_1^0_v1}
\end{align}
Let $g$ be a differentiable real valued function of $v_1,$ $ v_3$ and $v_5$ such that
\begin{equation}\label{eq:intro_g}
    \frac{\partial g}{\partial v_1} = \frac{1}{3} \alpha_2^0.
\end{equation}
Inserting definition \eqref{eq:intro_g} in equation \eqref{eq:alpha_1^0_v1} and integrating with respect to $v_1$ lead to:
\begin{equation}\label{eq:alpha10}
    \alpha_1^0 = 2 g - 3 v_1 \frac{\partial g}{\partial v_1} - 2 v_3 \frac{\partial g}{\partial v_3} - 3 v_4 \frac{\partial g}{\partial v_4}
\end{equation}
if the integration constant is set to zero. Thus far, we have successfully expressed $\alpha_1^0$ and $\alpha_2^0$ in terms of $g$. Let us proceed similarly with $\alpha_3^0$. Using again equations \eqref{potential_strict_ns_dern},
\begin{align}
  \alpha_3^0 &= -2 I_1^{\frac{3}{2}}\frac{\partial }{\partial I_1}\left(I_1^{-\frac{1}{2} }\alpha_3^0\right) + 2 I_1 \frac{\partial \alpha_3^0}{\partial I_1}\\[10pt]
  &= -2 I_1^{\frac{3}{2}} \frac{\partial }{\partial B_3}\left(\frac{1}{2} \alpha_1^0\right) + 2 I_1 \left(  \frac{\partial \alpha_3^0}{\partial v_1}\frac{\partial v_1}{\partial I_1} +  \frac{\partial \alpha_3^0}{\partial v_3}\frac{\partial v_3}{\partial I_1} + \frac{\partial \alpha_3^0}{\partial v_4}\frac{\partial v_4}{\partial I_1} \right)\\[10pt]
  &= -2 I_1^{\frac{3}{2}} \frac{\partial^2 \phi}{\partial I_1 \partial B_3} - 3 v_1 \frac{\partial \alpha_3^0}{\partial v_1} - 2 v_3 \frac{\partial \alpha_3^0}{\partial v_3} - 3 v_4 \frac{\partial \alpha_3^0}{\partial v_4}\\[10pt]
    &= - \frac{\partial \alpha_1^0}{\partial v_4}- 3 v_1 \frac{\partial \alpha_3^0}{\partial v_1} - 2 v_3 \frac{\partial \alpha_3^0}{\partial v_3} - 3 v_4 \frac{\partial \alpha_3^0}{\partial v_4}.
\end{align}
Using equation \eqref{eq:alpha10}, we obtain:
\begin{equation}
    \alpha_3^0 = \frac{\partial g}{\partial v_4} + 3 v_1 \frac{\partial }{\partial v_1}\left(\alpha_3^0 - \frac{\partial g}{\partial v_4}\right) + 2 v_3 \frac{\partial}{\partial v_3}\left(\alpha_3^0 - \frac{\partial g}{\partial v_4}\right) + 3 v_4 \frac{\partial}{\partial v_4}\left(\alpha_3^0 - \frac{\partial g}{\partial v_4}\right),
\end{equation}
yielding
\begin{equation}
    \alpha_3^0 = \frac{\partial g}{\partial v_4}.
\end{equation}
To sum up, $\alpha_i^0$ are zero for $i=4,5,6,7$ and
\begin{align}
  \alpha_1^0 &= 2g -3v_1\Dpu{g}{v_1} - 2 v_3\Dpu{g}{v_3} - 3 v_4\Dpu{g}{v_4}   \label{alpha_1_ns_p},\\
    \alpha_2^0 &= 3\Dpu{g}{v_1}\label{alpha_2_ns_p},\\
  \alpha_3^0 &= \Dpu{g}{v_4}.\label{alpha_3_ns_p}
\end{align}
This concludes the proof.
\end{proof}

\begin{remark}
To prove theorem \ref{thm:pot_inv}, we applied theorem \ref{thm:NS_invariant} to identify constraints -- namely, equations \eqref{potential_strict_ns} -- on the potential $\phi$ and derive the corresponding form of the subgrid tensor by sufficient conditions. This leads to the scalar function $g$ set by equation \eqref{eq:intro_g}. 
\end{remark}

\subsection{Positive total dissipation}\label{subsec:pot_tensor_positive}

So far, we showed that there exists a potential $\phi$ such that
\begin{equation}
  \phi_T = \tr\left(\frac{\partial(\phi_{visc}-\phi)}{\partial\S}\S\right)
\end{equation}
if the SGS model $\ttau$ has the form prescribed in theorem \ref{thm:pot_inv}. A sufficient condition for the dissipation $\phi_T$ to be positive is that the potential $(\phi_{visc}+\phi)$ is convex with respect to $\S$. In the next theorem, we express this condition in terms of $g$.

\begin{theorem}
Let $\ttau$ be a subgrid tensor such that its deviatoric part is
\begin{equation}\label{eq:pot_inv2}
  \begin{array}{ll} 
	\ttau^d &\displaystyle= \left(2g - 3 v_1 \frac{\partial g}{\partial v_1}-  2 v_3 \frac{\partial g}{\partial v_3}- 3  v_4 \frac{\partial g}{\partial v_4}\right)\S\\[20pt] 
	&\displaystyle+ \frac3{|\S|} \frac{\partial g}{\partial v_1}(\S^2)^d
	+\frac1{|\S|} \frac{\partial g}{\partial v_4}(\tOmega^2)^d.
  \end{array}
\end{equation}
We set $v^* = \frac{1}{\sqrt{6}}$. If the function $g$ is such that for all $v_1 \in [-v^*, v^*]$:
\begin{enumerate}
    \item the function $(v_3, v_4) \mapsto g(v_1, v_3, v_4)$ is convex,
    \item $g(v_1, 0, 0) \leq \nu,$
\end{enumerate}
then, $\ttau$ is invariant under the symmetry groups $G_t$, $Gal$, $SO(3)$, $G_p$ and $G_s$ of the Navier-Stokes equations and induces a positive total dissipation.
\end{theorem}

\begin{proof}
From equation \eqref{eq:pot_inv2}, the sum of the filtered viscous dissipation and the subgrid dissipation reads
\begin{align}
    \phi_T &= I_1 \left(2 \nu - 2g + 2 v_3 \frac{\partial g}{\partial v_3} + 3 v_4 \frac{\partial g}{\partial v_4} - v_4 \frac{\partial g}{\partial v_4}\right)\\
    &= 2 I_1 \left( \nu - g + v_3 \frac{\partial g}{\partial v_3} + v_4 \frac{\partial g}{\partial v_4}\right).
\end{align}
It follows from Lemma 2.9 of \cite{razafindralandy2005} that $v_1$ is a bounded function of $\tS$: $v_1 \in [-v^*, v^*]$. The function $g$ being convex with respect to the variables $(v_3, v_4)$,
\begin{equation}
  v_3 \frac{\partial g}{\partial v_3} + v_4 \frac{\partial g}{\partial v_4} \geq g - g(v_1,0,0),\qquad\text{for all }v_1 \in [-v^*, v^*].
\end{equation}
By assumption, $g(v_1, 0, 0) \leq \nu,$ so that $\phi_T \geq 0.$
\end{proof}

\section{Comparison with SGS tensors in the literature}\label{sec:ex_and_comp}

The first symmetry-preserving SGS model for the Navier-Stokes equations was developed in \cite{razafindralandy05d}. After the requirement of positive dissipation, the model in this article reads
\begin{equation}\label{eq:subgrid_dina}
    \ttau^d = \nu\left(2 g(v_1) - 3 v_1 \frac{\partial g}{\partial v_1}\right) \S + \frac{3 \nu}{| \S |} \frac{\partial g}{\partial v_1} (\S^2)^d. 
\end{equation}

This model is invariant not only under the symmetry groups $G_t$, $Gal$, $SO(3)$, $G_p$ and $G_s$, but also under transformations \eqref{eq:transformation_viscosity} involving the viscosity $\nu.$ This explains the explicit dependence on $\nu$ which is rather unusual in LES. As explained previously, we did not consider transformations \eqref{eq:transformation_viscosity}. The main difference between our construction and the model \eqref{eq:subgrid_dina} is the dependency of the model \eqref{eq:pot_inv2} on the filtered vorticity tensor $\tOmega$. Model \eqref{eq:subgrid_dina} is recovered from theorem \ref{thm:pot_inv} if $g$ is chosen to depend only on $v_1.$

Another SGS model which was developed with similar tools to ours is the model of Lund and Novikov \cite{lund1992}. It reads 

\begin{equation}
\label{eq:tau_ln}
\begin{array}{ll}
  \ttau^d &\displaystyle =  C_1| \S | \Delta^2 \S + C_2 \Delta^2 (\S^2)^d 
  + C_3 \Delta^2 (\tOmega^2)^d\\[10pt]
  &\displaystyle + C_4 \Delta^2 [\S, \tOmega]
  + \frac{C_5}{| \S |} \Delta^2  [ \S^2, \tOmega]
\end{array}
\end{equation}
where the $C_i,$ $1 \leq i \leq 5,$ are constants of the model, and $\Delta$ is the grid spacing. Kosovic's model \cite{kosovic_1997} is a particular case of \eqref{eq:tau_ln} in which $C_3=C_5=0$. Model \eqref{eq:tau_ln} can be compared to our model \eqref{thm:NS_invariant}  with $\alpha_4^0=\alpha_7^0 = 0$, namely
\begin{equation}
\label{eq:tau_part}
\begin{array}{ll}
\ttau^d &\displaystyle = \alpha_{1}^0 \S 
+ \frac{1}{| \S \| } \alpha_{2}^0 (\S^2)^d 
+ \frac{1}{|\S| } \alpha_{3}^0(\tOmega^2)^d\\[10pt]
&\displaystyle + \frac{1}{|\S| } \alpha_{5}^0 [\S, \tOmega]
+ \frac{1}{|\S|^2 } \alpha_{6}^0 [ \S^2,  \tOmega].
\end{array}
\end{equation}
Lund and Novikov model \eqref{eq:tau_ln} was built to be SO(3)-invariant, via the Cayley-Hamilton theorem. It is not invariant under the scaling group $G_s$. This explains the factor $|\S|$ difference between the two models. So, model \eqref{eq:tau_ln} cannot be a particular case of \eqref{eq:tau_part}. 

Also, our model \eqref{eq:pot_inv2} can be compared to the classical and well established Smagorinsky model
\begin{equation}
\ttau^d_{Smag} = \nu_t \S =  C \Delta^2 |\S| \S
\label{eq:smag}
\end{equation}
where $C$ is a constant and $\Delta$ is the computational grid size. The Smagorinsky model does not preserve the symetry group of the Navier-Stokes equations \cite{oberlack97,razafindralandy2005}. As such, the effective-viscosity term $\nu_t$ in Smagorinsky's model cannot be recovered from any function $g=g\left(v_1,v_3,v_4\right)$ since it depends explicitly on $|\S|$. However, our model retains only the leading term in $\S$ for the specific choice of $g=g\left(v_3\right)$ which leads to a family of SGS models that are colinear to the Smagorinsky model.

\section{Conclusion}

We described a formal procedure for deriving a family of SGS models that preserves the Lie symmetries of the incompressible and instantaneous Navier-Stokes equations. These models explicitly depend on the strain rate and vorticity tensors and account for both the deformation and the rotational parts of the filtered velocity field, enhancing their robustness and the possible range of applications. The obtained class of models is thus very general and different approaches can be considered in order to reduce the model complexity. The one chosen in this article is a stability condition through a positive total dissipation criterion. To this aim, the SGS models are required to derive from a convex potential, leading in turn to a simpler class of models where a single scalar function $g$ remains to be determined.

Determining and specifying $g$ is beyond the scope of this article. However, one may seek $g$ such that the model embeds well-known wall laws (for example, $\ttau\sim O(y^3)$ in near-wall regions, with $y$ being the normal-wall distance). This can also be a good starting point for machine learning techniques: it is currently an active field of research \cite{otto2025,kurz2025harnessing,girimaji2023turbulence,ling2016machine}.

Numerical simulations on model \eqref{eq:pot_inv2} are yet to be carried out. Nevertheless, closely related turbulence models built on symmetry preservation, but not incorporating the filtered vorticity tensor, have already undergone numerically validated. It is the case of the model \eqref{eq:subgrid_dina} which has been tested on the numerical simulation of a ventilated room in \cite{razafindralandy05d}. This model showed a good agreement with experimental data and outperformed both the classical and the dynamic Smagorinsky models, especially in near wall regions. It is also the case of the anisothermal model validated in \cite{razafindralandy2012lie}.

The methodology presented here is sufficiently general to enable the derivation of SGS turbulent models for more complex systems of equations. One particular and interesting application is the two-phase Navier-Stokes equations in its one-fluid formulation, for which there is a lack of robust turbulent models in the literature. There, the physics is more complicated than in the monophasic case and the number of SGS tensors to be modelled is higher. Nevertheless, the proposed methodology extends naturally. This provides a promising starting point for future work.

%\bibliographystyle{abbrv}
%\bibliography{Biblio}

\end{document}